\documentclass[aps,pre,twocolumn,superscriptaddress,floatfix,amsmath]{revtex4}

\usepackage{ae}
\usepackage{graphicx}
\usepackage{psfrag}
\usepackage{bm}
\usepackage{wasysym}
\usepackage{amssymb}

\begin{document}

\title{Geometry of phase separation}
\vskip 10pt
\author{Alberto Sicilia}
\affiliation{Universit\'e Pierre et Marie Curie -- Paris VI, LPTHE UMR 7589,
4 Place Jussieu,  75252 Paris Cedex 05, France}
\author{Yoann Sarrazin}
\affiliation{Universit\'e Pierre et Marie Curie -- Paris VI, LPTHE UMR 7589,
4 Place Jussieu,  75252 Paris Cedex 05, France}
\author{Jeferson J. Arenzon}
\affiliation{Instituto de F\'\i sica, Universidade Federal do 
Rio Grande do Sul, CP 15051, 91501-970 Porto Alegre RS, Brazil} 
\author{Alan J. Bray}
\affiliation{School of Physics and Astronomy, University of Manchester, 
Manchester M13 9PL, UK}
\author{Leticia F. Cugliandolo}
\affiliation{Universit\'e Pierre et Marie Curie -- Paris VI, LPTHE UMR 7589,
4 Place Jussieu,  75252 Paris Cedex 05, France}

\begin{abstract}
  We study the domain geometry during spinodal decomposition of a
  $50$:$50$ binary mixture in two dimensions.  Extending arguments
  developed to treat non-conserved coarsening, we obtain
  $approximate$ analytic results for the distribution of domain areas
  and perimeters during the dynamics.  The main approximation is to
  regard the interfaces separating domains as moving independently.
  While this is true in the non-conserved case, it is not in the
  conserved one. Our results can therefore be considered as a `first
  order' approximation for the distributions.
  In contrast to the celebrated Lifshitz-Slyozov-Wagner distribution
  of structures of the minority phase in the limit of very small
  concentration, the distribution of domain areas in the $50$:$50$
  case does not have a cut-off.  Large structures (areas or
  perimeters) retain the morphology of a percolative or critical
  initial condition, for quenches from high temperatures or the
  critical point respectively. The corresponding distributions are
  described by a $c A^{-\tau}$ tail, where $c$ and $\tau$ are exactly
  known. With increasing time, small structures tend to have a
  spherical shape with a smooth surface before evaporating by
  diffusion.  In this regime the number density of domains with area
  $A$ scales as $A^{1/2}$, as in the Lifshitz-Slyozov-Wagner theory.
  The threshold between the small and large regimes is determined by
  the characteristic area, ${\rm A} \sim [\lambda(T)
  t]^{2/3}$. Finally, we study the relation between perimeters
  and areas and the distribution of boundary lengths, finding results
  that are consistent with the ones summarized above. We test our
  predictions with Monte Carlo simulations of the $2d$ Ising Model.
\end{abstract}

\maketitle

\section{Introduction}


Phase separation is the process whereby a binary mixture of components
$A$ and  $B$, initially in a  homogeneous phase, demix  leading to the
coexistence of two phases, one rich  in $A$ and the other in $B$.  The
system, initially  in an unstable spatially uniform  state, performs a
coarsening   process   to   approach  its   thermodynamically   stable
phase-separated state~\cite{BrayReview}.

In the  case of fluids, hydrodynamic  effects may be  important to the
demixing process  \cite{Siggia}; these interactions are  hard to treat
analytically and  the results  of numerical simulations  are sometimes
controversial.  Some  studies even  claim  that  dynamical scaling  is
broken by hydrodynamic transport  \cite{Yeomans}.  In the following we
focus   on   the  phase   separation   process   in  systems   without
hydrodynamics.  A typical  realization is  phase separation  in binary
alloys, high-viscosity fluids and polymer blends.


The time-dependent order parameter characterising the phase separation
phenomenon  is  a  continuous  scalar field,  $\phi(\vec  r,t)$,  that
represents  the local difference  in concentration  of the  two phases
normalized by the sum of the averaged concentrations. Its evolution is
described by a phenomenological Langevin-like equation:
\begin{equation}
\frac{\partial{\phi}}{\partial{t}} = 
\overrightarrow{\triangledown}\cdot\left[ M(\phi) 
\overrightarrow{\triangledown}
 \left(\frac{\delta H [\phi]}{\delta \phi}\right) \right] 
+ \eta
\; .
\label{eq:Langevin} 
\end{equation}
$H[\phi]$ is a Ginzburg-Landau type `free-energy' with an elastic term
and a double-well potential.  The stochastic field $
\eta(\vec r,t)$ represents a conserved thermal noise, while the
mobility $M(\phi)$ is in general a function of the field $\phi$, as
indicated.  When $M(\phi)=1$, Eq.~(\ref{eq:Langevin}) is known as the
Cahn-Hilliard equation \cite{Cahn-Hilliard}, or model $B$ in the
Hohenberg-Halperin classification of critical dynamics~\cite{HH}.
Several  discrete  models to  study  the  phase-separation process  in
binary   alloys  have   also  been   proposed  and   studied   in  the
literature. These  are lattice gases  that are themselves  mapped onto
Ising       models       with       locally      conserved       order
parameter~\cite{Kawasaki,Lebowitz}.   The dynamics  follow stochastic
rules for the  interchange of nearest neighbour $A$  and $B$ molecules
or,  in the  spin language,  the reversal  of a  pair  of neighbouring
antiparallel   spins. 


Two microscopic  processes contribute to phase ordering  dynamics with
locally conserved order parameter, namely {\it bulk} and {\it surface}
diffusion. In  bulk diffusion a molecule  separates (evaporation) from
the surface  of a domain,  diffuses within the neighbouring  domain of
the opposite  phase, and  finally attaches to  its original  domain or
another one. In the context of  an Ising model simulation, there is an
activation energy for this process and one can check that the dominant
growth mechanism  is the transport of material through the  bulk from
domain boundaries with {\it large} curvature to domain boundaries with
{\it small}  curvature. In surface  diffusion molecules `walk'  on the
interface at  no energy  cost. This mechanism  leads to the  motion of
whole domains  in the sample and  thus the possibility  of merging two
domains together when they collide. In the usual Kawasaki~\cite{Kawasaki}
spin-exchange dynamics or in model B's evolution~\cite{Gemmert}, the 
early dynamics is surface diffusion driven
and at later  time the dominant process becomes  bulk diffusion.  There
exist, however, modified versions where either bulk or surface 
diffusion are suppressed~\cite{Gemmert,Puri-Bray-Lebowitz,Barkema}.

For deep temperature quenches, $M(\phi) \rightarrow 0$, bulk diffusion is effectively eliminated,
and domain growth proceeds by surface motion. In quenches to moderate 
subcritical temperatures, on the other hand, the mobility does not play 
an important  role, $M(\phi)\simeq \mbox{const.}$, and the domain growth
is  bulk-driven.


%


Let us now review in some  detail the main features of coarsening with
locally conserved  order parameter.  Using  the Cahn-Hilliard equation,
it  can be  easily shown  that the  radius, $R(t)$,  of a  {\it single
spherical} domain of  negative phase ($\phi = -1$)  in an infinite sea
of positive phase ($\phi=+1$), evolves from time $t=0$ to time $t$ as
\begin{equation}
 R^3(t) = R^3(0) - \frac{3}{2} \sigma t.
\label{eq:t13}
\end{equation}
with  $\sigma$ a  parameter that  quantifies the  surface  tension.  A
domain with  initial radius  $R_i$ thus evaporates  in a time  $t \sim
R_i^3$ in  contrast to the  non-conserved order parameter  dynamics in
which the  area within a  boundary simply shrinks under  the curvature
force in a time $t \sim R_i^2$~\cite{AC}.

Lifshitz-Slyozov~\cite{Lifshitz} and Wagner~\cite{Wagner} studied, for
a {\it three-dimensional} system, the growth and shrinkage of domains of one
phase embedded in one large domain  of the other phase in the limit of
small  minority  phase concentration,  $c  \rightarrow  0$.  In  their
celebrated papers they realized that  domain growth at the late stages
is limited  by matter diffusion  through the majority domain.  In this
case the evolution of a domain of the minority phase with radius $R_i$
immersed in a sea of  the majority phase that is `supersaturated' with
the  dissolved minority species~\cite{Lifshitz,BrayReview}  can follow
two  paths: the  domain evaporates  by diffusion  if $R_i<R_c$,  or it
grows  by absorbing  material from  the majority  phase  if $R_i>R_c$,
where  $R_c$  is a  time-dependent  `critical  radius'. This  critical
radius turns  out to  be the only  characteristic length-scale  in the
system, ${\rm  R}(t)$, and serves to scale  all correlation functions
according  to  the dynamic  scaling  hypothesis.  It  grows as  ${\rm
R}(t)=R_c(t)\sim t^{1/3}$.

Lifshitz-Slyozov~\cite{Lifshitz}  also derived  an expression  for the
density of droplets of the minority phase with linear size $R$ in $d=3$.  Three
important properties of the Lifshitz-Slyozov distribution are:
\begin{itemize}
\item
The  distribution  of droplet  radii  has  an  upper cut-off,  $R_{\rm
max}(t)$, where $R_{\rm max}(t) \sim  t^{1/3}$ is a constant, equal to
3/2, times the critical radius $R_c(t)$.
\item
The decay close to the cut-off is exponential.
\item
The density of small objects, $R\sim 0$, satisfies scaling and behaves
as  $n(R,t) \sim R_c(t)^{-4}\,(R/R_c(t))^2$,  where $n(R,t)dR$  is the
number of  droplets per unit volume  with radius in  the interval $(R,
R+dR)$.
\end{itemize}
Later simulations established that the scaling functions depend on the
minority concentration~\cite{Puri88,Toral}.

The Lifshitz-Slyozov  calculation can easily be extended  to any space
dimension $d>2$ \cite{Yao}. However, the  limit $d \to 2$ is singular,
and  does not  commute with  the  limit $c  \to 0$.  Rogers and  Desai
\cite{RogersDesai} showed, however, that the usual scaling forms apply
in  $d=2$, with  ${\rm R}(t) \sim  t^{1/3}$, for large $t$ at small 
non-zero volume fraction $c$.

More recently, Huse used scaling and energetic arguments to generalize
the Lifshitz-Slyozov growth law and argued that it should also apply
to critical quenches with equal volume fractions of the two
phases~\cite{Huse}.  Numerical simulations~\cite{Huse,Amar}, in 
agreement also with the
earlier ones in \cite{Lebowitz}, suggest that the typical domain
radius scales in time as ${\rm R}(t) \sim t^{1/3}$ for any value of
$c$, even in the 50:50 case. The scaling function for the distribution
of domain areas has not been analyzed in this case.


If we take also into  account the competition between bulk and surface
diffusion  the growth  law  is  modified at  early  times. The  former
process is the  one responsible for the scaling  of global observables
with a {\it typical} domain  length ${\rm R}(t)\sim t^{1/3}$ while the
latter  yields a  slower time-dependence,  ${\rm R}(t)  \sim t^{1/4}$,
that is  important only  at relatively short  times after  the 
quench~\cite{Mazenko,Corberi,Szabo}.
The  temperature-dependent  crossover can  be  seen,  for example,  in
numerical  simulations  with  Kawasaki  dynamics~\cite{Gemmert}. The  
crossover  time
diverges when  $T\to 0$.   This observation has  been used  to develop
accelerated algorithms to simulate  discrete models in which only bulk
diffusion   processes   are    considered,   which   should   describe
phase-separation correctly at late times after the quench.

Phase-separation in the  Kawasaki spin-exchange dynamics is equivalent
to a  Cahn-Hilliard equation with order  parameter dependent mobility.
In~\cite{Puri-Bray-Lebowitz},   a   model   with  $M(\phi)=1-   \alpha
\phi^{2}$ was  studied.  The time dependent  structure factor exhibits
dynamical   scaling,   and  the   scaling   function  is   numerically
indistinguishable from the Cahn-Hilliard one, consistent with what was
expected from numerical studies with Kawasaki dynamics.



In  this  paper  we  study  the morphology  of  domain  and  perimeter
structures in  the spinodal decomposition of a  two dimensional system
with {\it equal  concentrations} of the two phases.   In particular, we
analyse  the distributions of  the domain  areas and  their associated
perimeters, and  the relation between areas and  perimeters during the
evolution.   We  consider  bulk-and-surface  diffusion and  just-bulk
diffusion processes.

Extending  a formalism previously  developed for  the study  of domain
growth  in  the  non-conserved case  \cite{us,Alberto-PRE,us-EPL},  we
propose an analytic form for  the domain size distribution function in
its full-range  of variation, and we  test it with  simulations on the
two-dimensional Ising model ($2d$IM).  Our analytic prediction for the
distribution  of small  areas is  the result  of one  hypothesis: that
interfaces move  independently.  This assumption is  valid for domains
of any size in curvature driven  dynamics: the fission of a big domain
into two  smaller ones  or the  coalescence of two  domains to  form a
bigger    one   are   forbidden    in   the    continuous   Allen-Cahn
description~\cite{AC}, and are not important in the heat-bath dynamics
of  the  Ising  model  with  non-conserved order  parameter.   In  the
conserved  order parameter  case, even  with $c  \rightarrow  0$, this
assumption does not strictly  hold and corrections must, in principle,
be included. Indeed,  already Lifshitz and Slyozov made  an attempt to
go beyond their simple model and account for coalescence when $c\to 0$.
Later,  it became  clear  that in  locally  conserved order  parameter
dynamics,  the   dominant  effect  not accounted  for  in  the  simple
description  that takes domain-boundaries  as independent  objects was
interdomain correlations rather than coalescence.  For a discussion of
the limit $c \rightarrow  0$ considering interactions between droplets
see~\cite{Marder}.

Therefore,   our  analytical   results  are   just  a   `first  order'
approximation.  Still, as we shall see  in the main body of the paper,
this approximation  yields a very  good description of  numerical data
obtained  with Monte  Carlo simulations.  The quantities  on  which we
focus are domain areas and `hull enclosed areas', where the latter are
the areas enclosed within the outer boundaries (or `hulls') of domains,
and  domain  perimeters.  The  distribution  functions  for  the  hull
enclosed and  domain areas are  $n_{h,d}(A,t)$, where $n_{h,d}(A,t)dA$
is the number of hulls (h) or domains (d) per unit area of the system 
with area in the range $(A,A+dA)$.

The main  properties  of these distributions are:



\begin{itemize}
\item The number density of domain and hull-enclosed areas satisfy
  scaling:  $n_{h,d}(A, t) = t^{-4/3} f_{h,d} (A/t^{2/3})$. The argument
  of the scaling functions arises from the fact that the characteristic
  area of hulls and domains grows as $t^{2/3}$. The prefactor
  $t^{-4/3}$ follows from the fact that there is of order one domain
  per scale area. 
\item The scaling functions, $f_{h,d}(x)$, do not have any cut-off and
  extend to infinite values of $x$ falling-off as $(2) c_{h,d}
  x^{-\tau}$, with $x=A/(\lambda_{h,d} t)^{2/3}$. The constants $c_{h,d}$ 
  are the ones in the initial (or quasi-initial --  see below) area 
  distribution, $n_{h,d}(A_i) \sim (2)c_{h,d}/A_i^\tau$, 
  for $A_i \to \infty$. The prefactor $c_h$ is known exactly,  
  $c_h=1/8\pi\sqrt{3}$~\cite{Cardy}.  The factor 2 is present when
  quenching from high temperature, $T_0>T_c$, and  is absent when the
  initial condition is the critical Ising one, $T_0=T_c$.  The exponent
  $\tau$ depends on whether we consider hull-enclosed ($\tau=2)$ or
  domain areas, and in the latter case on the initial condition, {\it
    i.e.} infinite or critical temperature. In both cases it is very
  close to $2$.
\item After a quench from high temperature to a sufficiently low
  working temperature, $T \ll T_c$, the small-argument behaviour of the
  scaling function is $f(x) \propto \sqrt{x}$, in agreement with the 
  Lifshitz-Slyozov-Wagner prediction for the small concentration limit. 
  At higher working temperature and for critical initial conditions 
  the behaviour is modified in a way that we describe in the text.
\end{itemize}

The  paper is  organised as  follows.  In  Sect.~\ref{sec:analytic} we
describe an approximate analytic  derivation of the time-dependent hull
enclosed and domain area distributions. These arguments do not rely on
any scaling hypothesis but rather support its validity. We compare our
approach      to      the      celebrated      Lifshitz-Slyozov-Wagner
theory~\cite{Lifshitz,Wagner}. In  Sect.~\ref{sec:numeric} we show our
numerical results for  the statistics of areas in  the $2d$IM evolving
with locally conserved dynamics. We use variants~\cite{Barkema} of the
Kawasaki  algorithm~\cite{Kawasaki} that  we briefly  explain  in this
Section.  The use of different  algorithms allows us to switch surface
diffusion on  and off, and  pinpoint the relative importance  of these
processes.  Section~\ref{sec:domain-walls} is devoted to the analysis,
both analytical and numerical, of  the geometry of domain walls during
the dynamics and their  relation to the corresponding areas.  Finally,
in  the Conclusion  we summarize  our  results and  we discuss  future
studies along these lines.

\section{Statistics of areas: analytic results}
\label{sec:analytic}

In this Section we analyze the number density of hull-enclosed and
domain areas. A domain is a region of connected aligned spins. 
Each domain has one external perimeter which is called the hull. The
hull-enclosed area is the total area within this perimeter, {\it i.e.} the
domain area plus the area of any internal sub-domains.

\subsection{Initial distribution}
\label{subsec:initial}

We study the coarsening dynamics after a quench from $T_0\to\infty$
and $T_0=T_c$.
The first case is very similar to critical percolation,
as we explained in \cite{Alberto-PRE}. Equilibrium infinite temperature 
initial conditions (fraction of up spins $=1/2$) are below the
critical random percolation point ($p_c \approx 0.59$) for a square 
lattice in $d=2$. After a few MC steps, however, the system reaches the 
critical percolation condition, e.g.\ the expected $A^{-2}$ tail is observed 
in the distribution of hull enclosed areas. 
We have checked that this is 
so from the analysis of several correlation functions as well as the 
distribution of structures. The same happens for non-conserved order 
parameter dynamics~\cite{Alberto-PRE}. This fact justifies the use of the
Cardy-Ziff exact result for the distribution of hull-enclosed areas at
critical percolation~\cite{Cardy} as our effective initial condition
from $T_0\to\infty$. From a more general perspective, one can argue 
that if the system is coarse-grained to the scale ${\rm R}(t)$, and 
${\rm R}(t)$ is large compared to the lattice spacing, a continuum 
description becomes appropriate, for which the percolation threshold 
is precisely $1/2$. At the scale ${\rm R}(t)$, the system is disordered 
and has the character of the percolation model at threshold. 
 
In Section II in \cite{Alberto-PRE} we recalled the equilibrium
distribution of hull-enclosed and domain areas, domain walls and their
geometrical relation to their associated areas at critical percolation
and critical Ising initial conditions. In Section III we derived some
generic results that follow from the use of sum rules and the scaling
hypothesis. We do not repeat the description of these properties here
but we refer the reader to \cite{Alberto-PRE} for further details.

For future reference we simply list here the initial distributions, 
$n_{h,d}(A_i,t_i)$, of hull-enclosed and domain areas 
\cite{Cardy,Stauffer,Stella}:
\begin{align}
n_h(A_i,t_i) &=\frac{(2)c_h}{A_i^2} \nonumber
\\
n_d(A_i,t_i) &= \frac{c_dA_0^{\tau-2}}{A_i^\tau}
\; , & \qquad \tau = \frac{379}{187} 
\label{eq:initial-domains-Tc}
\\
n_d(A_i,t_i) &= \frac{2c_dA_0^{\tau'-2}}{A_i^{\tau'}}
\; , & \qquad \tau' = \frac{187}{91} \nonumber
\end{align}
where the exponent $\tau$ describes the critical Ising state  
and $\tau'$ the critical percolation state. 

The  expression for  $n_h(A_i,t_i)$ is  asymptotically exact  for areas
$A_i$ large compared to the microscopic area, $A_0 \approx a^2$, where
$a$  is the  lattice spacing.  The factor  2 is  present  for critical
percolation  but  absent  for  the critical  Ising  initial  condition
\cite{Cardy}.  In the former case, the ``initial'' time $t_i$ is taken
to be a few MC steps, such that the system  has effectively reached the
percolative critical state.
The constant $c_h$ is known analytically, $c_h=1/8\pi\sqrt{3}$~\cite{Cardy}.

The expressions  for $n_d(A_i,t_i)$ require a  little more discussion.
The key  point is that the  constant $c_h$ is very  small ($c_h \simeq
0.023$),   so   that  ``domains-within-domains''   are   rare.  As   a
consequence,  the domain area  distributions are  very similar  to the
corresponding  distributions  of  hull  enclosed  areas.  However,  as
discussed   at   length  in   \cite{Alberto-PRE},   the  domain   area
distributions must  fall off  faster than $A^{-2}$  in order  that the
total domain area,  per unit area of the system,  be finite (and equal
to unity).  The smallness  of $c_h$ suggests  that they will  fall off
only slightly  faster than $A^{-2}$,  and indeed the  exponents $\tau$
and  $\tau'$ in  Eq.~(\ref{eq:initial-domains-Tc}) take  the numerical
values  $\tau \simeq  2.027$ and  $\tau' \simeq  2.055$.  Formally, if
$c_h$  is regarded as  a small  parameter, one  finds $\tau=2+{\cal O}(c_h)$,
$\tau'   =   2+  {\cal O}(c_h)$   and   
$c_d   =   c_h  +   {\cal O}(c_h^2)$~\cite{Alberto-PRE}. This line  
of reasoning also accounts for  the factor 2
that  (again  to leading  order  in $c_h$)  appears  in  the third  of
Eqs.~(\ref{eq:initial-domains-Tc}) but not the second.


\subsection{Characteristic domain length}

It is by now well-established~\cite{Lifshitz,Huse} 
that the spatial equal-time correlation function in demixing systems 
is correctly described by the dynamic scaling hypothesis with a characteristic
length ${\rm R}(t)\sim t^{1/3}$ in agreement with the Lifshitz-Slyozov-Wagner 
prediction~\cite{Lifshitz,Wagner} and the extension beyond the small 
concentration limit derived by Huse~\cite{Huse}. 
The effect of temperature fluctuations is 
expected to be described by a $T$-dependent prefactor,
${\rm R}(t,T)=[\lambda(T) t]^{1/3}$.  In the following we do not 
write the $T$ dependence explicitly.

\subsection{Large structures}
\label{subsec:large-areas}

It is natural to assume that at time $t$, as for non-conserved order
parameter dynamics, large structures, characterized
by a long linear dimension $R \gg {\rm R}(t)$, have not changed much
with respect to the initial condition. We shall support this claim
with the numerical results. Thus, for sufficiently large hull-enclosed
areas such that the time-dependence can be neglected we expect
\begin{eqnarray}
n_h(A,t) 
&\approx& 
\frac{(2)c_h}
{
A^2
} 
\; , 
\qquad
A \gg t^{2/3}
\; . 
\label{eq:large-hulls}
\end{eqnarray}

Similarly, for large domains, the area dependence of their distribution follows that of the initial condition, 
Eq.~(\ref{eq:initial-domains-Tc}).


\subsection{Small structures}
\label{subsec:small-areas}

Small structures, such that $R\ll {\rm R}(t)$, are mostly embedded in
very large domains. To a first approximation we shall assume that they are 
{\it independent}. Moreover,  they are not 
expected to have holes of the opposite phase within, implying the 
equivalence between hull-enclosed and domain areas at these scales. 
Indeed, any smaller structure placed within must have evaporated by time $t$.
We then propose that the number density of {\it small} hull-enclosed 
or domain areas at time $t$ can be written as a function of the initial
distribution,
\begin{equation}
n(A,t) \approx \int_0^\infty dA_i \; \delta(A-A(t,A_i)) \, n(A_i,t_i) 
\; ,
\label{eq:integration} 
\end{equation}
with $A_i$ the initial area and $n(A_i,t_i)$ their number density
at the initial time $t_i$. $A(t,A_i)$ is the area of a domain at
time $t$ that had area $A_i$ at time $t_i$.  In writing
this equation we have implicitly assumed that an area cannot
split into two, and that two such areas cannot coalesce, which is not 
strictly true in conserved order parameter dynamics.

Note that for sufficiently large areas so that the time-dependence 
is not important and $A\sim A_i$ Eq.~(\ref{eq:integration}) immediately
yields $n(A,t) = n(A_i,t_i)$ and Eq.~(\ref{eq:large-hulls}) is recovered.
It has to be stressed, however, that Eq.~(\ref{eq:integration}) 
does not strictly apply in this case.
 
Assuming that the small areas are {\it circular}
\begin{equation}
A^{3/2}(t,A_i) = A_i^{3/2} - \lambda_h(T) (t-t_i) 
\; , 
\end{equation}
see Eq.~(\ref{eq:t13}), and after a straightforward calculation using
Eq.~(\ref{eq:initial-domains-Tc}) for the initial distribution one finds
\begin{equation}
(\lambda_h t)^{4/3} \; n_h(A,t) 
= 
\frac{ (2)c_h \ 
\left[\displaystyle \frac{A}{(\lambda_h t)^{2/3}} \right]^{1/2} }
{
\left\{
1+ \left[\displaystyle \frac{A}{(\lambda_h t)^{2/3}} \right]^{3/2}
\right\}^{5/3}
} 
\label{eq:nh-allA}
\end{equation}
for hull-enclosed areas.  This prediction has the expected scaling
form $n_h(A,t) = t^{-4/3} f(A/t^{2/3})$ corresponding to a system with
characteristic area ${\rm A}(t) \sim t^{2/3}$ or characteristic
length scale ${\rm R}(t) \sim t^{1/3}$.  At very small areas,
$A\ll (\lambda_h t)^{2/3}$, where our approximations are better
justified, one has
\begin{equation}
(\lambda_h t)^{4/3} \; n_h(A,t) 
\approx
(2)c_h \;
\left[ \frac{A}{(\lambda_h t)^{2/3}} \right]^{1/2}
\; . 
\label{eq:nh}
\end{equation}
As expected, taking the limit $A/(\lambda_ht)^{2/3} \gg 1$
in Eq.~(\ref{eq:nh-allA}) one recovers Eq.~(\ref{eq:large-hulls}). Although 
this limit goes beyond the limit of validity of Eq.~(\ref{eq:nh-allA}) 
we shall propose that Eq.~(\ref{eq:nh})  actually holds, at least 
approximately, for all values of $A/(\lambda_h t)^{2/3}$. 
 
In \cite{Alberto-PRE} we studied non-conserved order parameter
dynamics and we derived the number density of domain areas from the
one of hull-enclosed areas. The key fact in this case was that we
could treat the distribution of hull-enclosed areas {\it exactly} and
then use a small $c_h \simeq 0.023$ expansion to get the statistical
properties of domains. In the case of phase separation our results for
hull-enclosed areas are already approximate. Still, the relation
between hull-enclosed area distribution and domain area distribution
obtained in \cite{Alberto-PRE} should remain approximately true, as a
first order expansion in small $c_h$, since small domains are not
expected to have structures within.  In conclusion, for 
large areas and long times such that a regularizing 
microscopic area $A_0=\lambda_d t_0$ can be neglected,  
we expect the same functional form as the one given in
Eq.~(\ref{eq:nh-allA}) with $c_d = c_h + {\cal O}(c_h^2)$, $\lambda_d =
\lambda_h[1 + {\cal O}(c_h)]$ and the power $5/3$ in the denominator replaced by
$(2\tau+1)/3$ or $(2\tau'+1)/3$:
\begin{eqnarray}
(\lambda_d t)^{4/3}  n_d(A,t) &\simeq&  
\frac{(2) c_d  \; 
\left[\displaystyle \frac{A}{(\lambda_d t)^{2/3}} \right]^{1/2}}
{\left\{1+ 
\left[\displaystyle \frac{A}{(\lambda_d t)^{2/3}} \right]^{3/2} 
\right\}^{(2\tau+1)/3}}
\nonumber\\
&\equiv& 
g(x)
\label{guess1confirmed}
\end{eqnarray}
with $x=A/(\lambda_dt)^{2/3}$, 
and 
\begin{equation}
(\lambda_d t)^{4/3}  n_d(A,t) \simeq  
(2) c_d  \; 
\left[ \frac{A}{(\lambda_d t)^{2/3}} \right]^{1/2}
\label{guess1confirmed-1}
\end{equation}
for $A\ll (\lambda_d t)^{2/3}$. This expression can be compared to 
Eq.~(49) in \cite{Alberto-PRE} valid for non-conserved order parameter 
dynamics.

Let  us  emphasize again  the  main  approximation  of our  analytical
approach:  we  are considering  each  domain  area  as an  independent
entity. This is strictly  true for the non-conserved order parameter
\cite{us,  Alberto-PRE} but is  an approximation  in the  conserved
order parameter problem.

\subsection{Super universality}
\label{subsec:small-areas}

All the results above are valid for the bulk diffusion driven case.
What happens if we consider the case in which bulk and surface
diffusion are in competition, or whether we include quenched disorder
in the couplings?
If we suppose that all these systems belong to the same dynamical
universality class, the scaling functions being the same, then
Eq.~(\ref{eq:nh-allA}) can be generalized in the form
\begin{equation}
{\rm R}^4(t) \; n_h(A,t) 
= 
\frac{ (2)c_h \ 
\left[\displaystyle \frac{A}{{\rm R}^2(t)} \right]^{1/2} }
{
\left\{
1+ \left[\displaystyle \frac{A}{{\rm R}^2(t)} \right]^{3/2}
\right\}^{5/3}
} 
\; 
\label{eq:nh-allA-super}
\end{equation}
and similarly for the domain area distribution.  The time-dependence
in ${\rm R}(t)$ should include all regimes ({\it e.g.}  $t^{1/4}$ and
$t^{1/3}$ in the clean case with surface and bulk diffusion) and can
be extracted from the dynamic scaling analysis of the correlation
functions.  We shall check numerically the super-universality
hypothesis.

\section{Statistics of areas: numerical tests}
\label{sec:numeric}

To test our analytic results we carried out numerical simulations on
the $2d$ square-lattice Ising model ($2d$IM) with periodic boundary
conditions.

\begin{equation}
H = - \sum_{\langle i,j\rangle} J_{ij} \sigma_i \sigma_j
\label{model}
\end{equation}

where $\sigma_i= \pm 1$. We will start considering the pure model,
$J_{ij}=J>0$ $\forall ij$, and then the random bond model in which the $J_{ij}$ 
are random variables uniformly distributed over the
interval $[1/2, 3/2]$.
We used several versions of conserved order parameter dynamics that
switch on and off surface diffusion. These are Kawasaki dynamics at
finite temperature including both surface and bulk diffusion, and
accelerated bulk diffusion in which surface diffusion is totally
suppressed. Bulk diffusion needs to overcome energy barriers; thus
this variant runs at finite temperature only. In all cases we
implemented the continuous time method and the algorithms become
rejection free.  A detailed description of these algorithms
appeared in~\cite{Barkema}.  Domain areas are identified with the
Hoshen-Kopelman algorithm~\cite{Hoshen}.

\begin{figure}[h]
\includegraphics[width=4.cm]{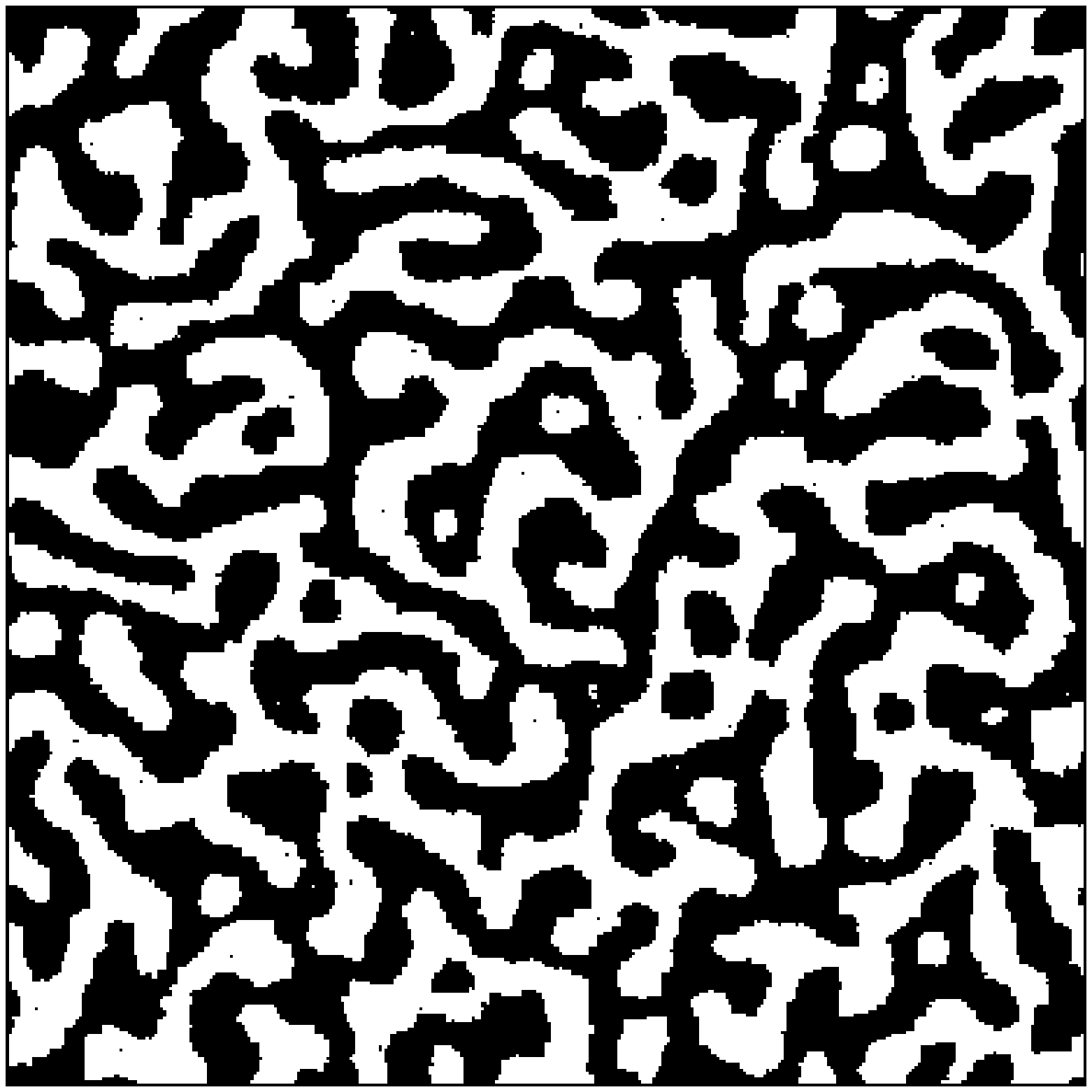}
\includegraphics[width=4.cm]{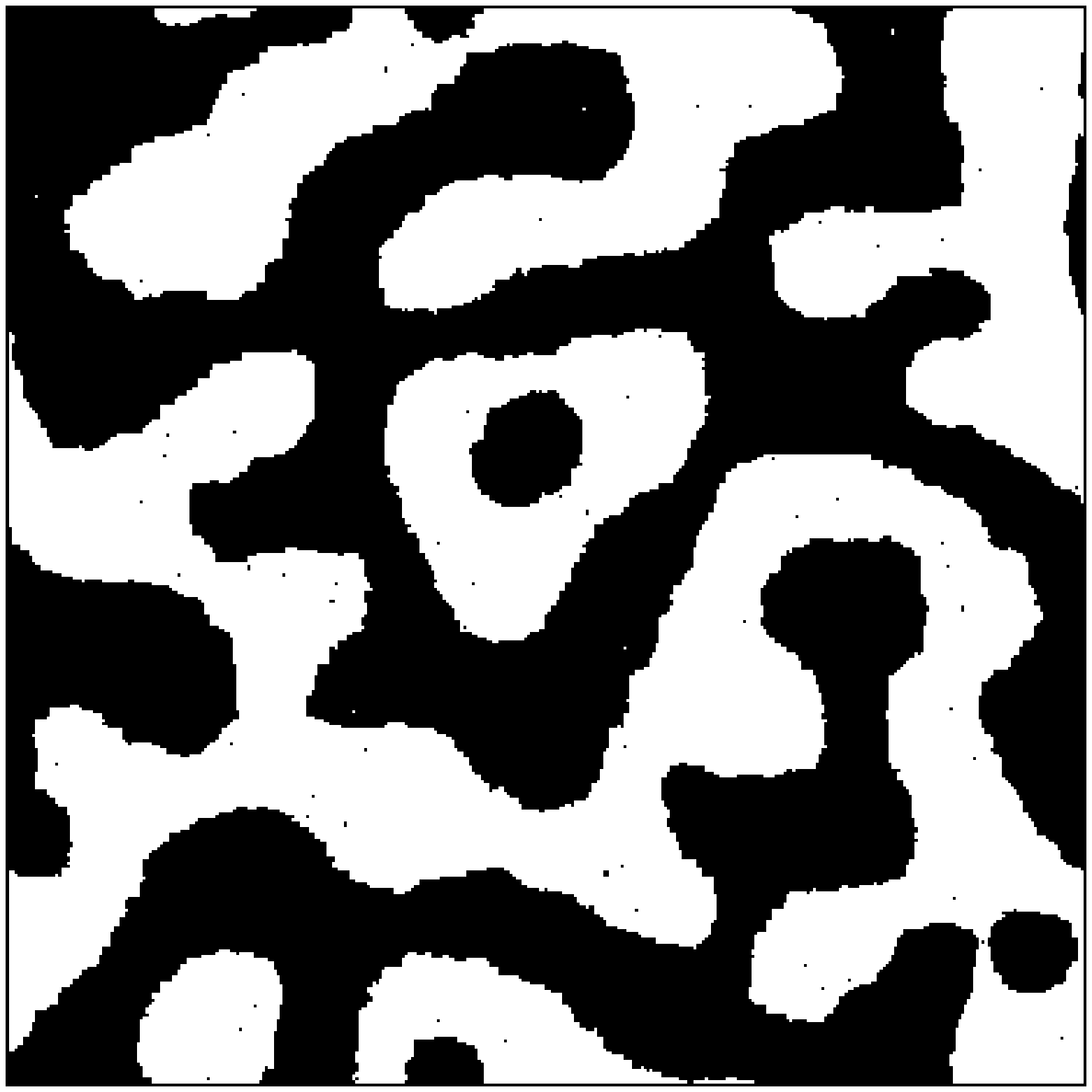}
\caption{Snapshots of the $2d$ Ising model evolving with locally
  conserved order parameter at $t=1000$ and $t=16000$ equivalent MC
  steps using the accelerated bulk algorithm explained in
  Sect.~\ref{sec:numeric}. The initial condition is a random configuration 
  of $\pm 1$ spins taken with probability one half, that is to say an infinite 
  temperature state. Evolution occurs at $T=1.0$.}
\label{fig:snapshots0}
\end{figure}

All data have been obtained using systems with size $L^2=10^3\times
10^3$ and $10^3$ runs using independent initial conditions.
We ran at different temperatures specified below.  We considered two
types of initial conditions, equilibrium at infinite temperature,
$T_0\to\infty$, and equilibrium at the critical Ising temperature,
$T_0=T_c$. All results are presented in equivalent Monte Carlo (MC)
times.  An example, for $T_0\to\infty$, is presented in 
figure~\ref{fig:snapshots0}, with snapshots taken at different times. 

\subsection{Domain areas}
\label{subsec:coarseningT0}

In this Section we show data
obtained with the accelerated bulk diffusion algorithm but we checked
that similar results are obtained with finite-temperature
Kawasaki dynamics.

\subsubsection{Infinite temperature initial condition, low $T$ evolution}
\label{subsubsec:Tinfty}

\begin{figure}[h]
\includegraphics[width=225pt]{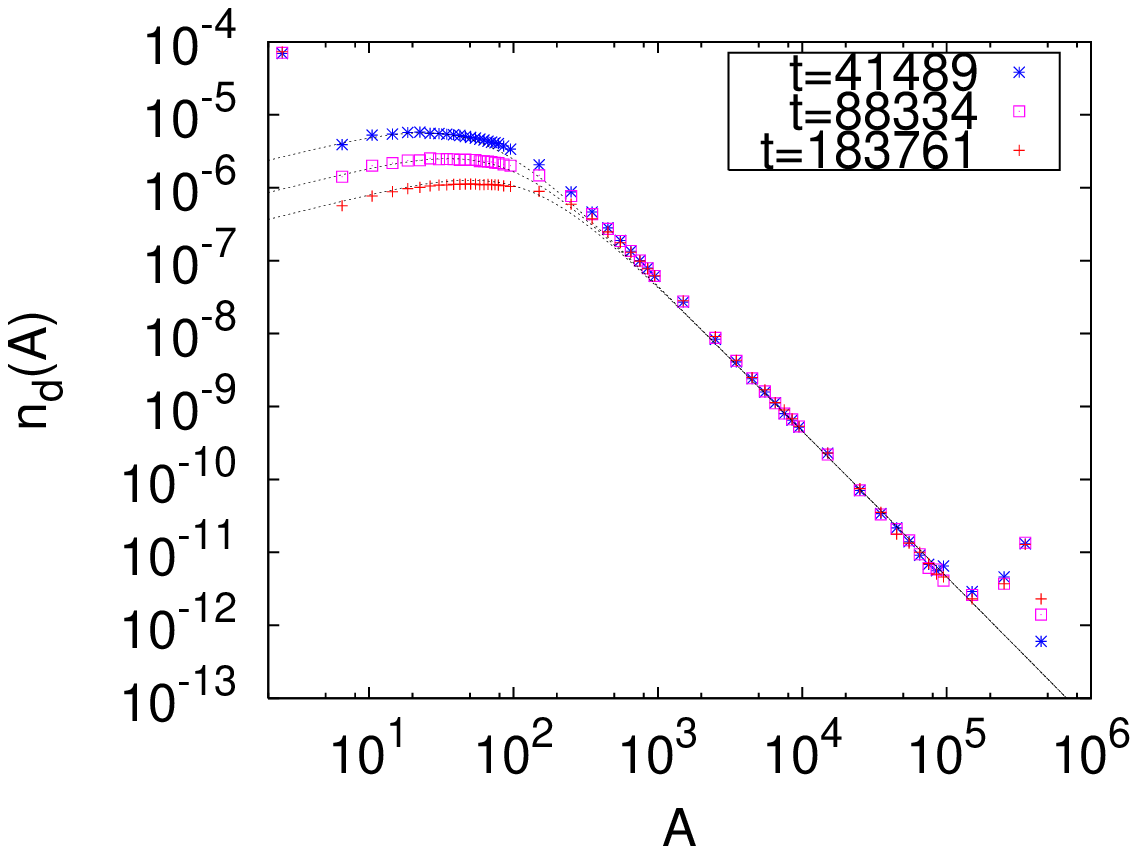}
\includegraphics[width=225pt]{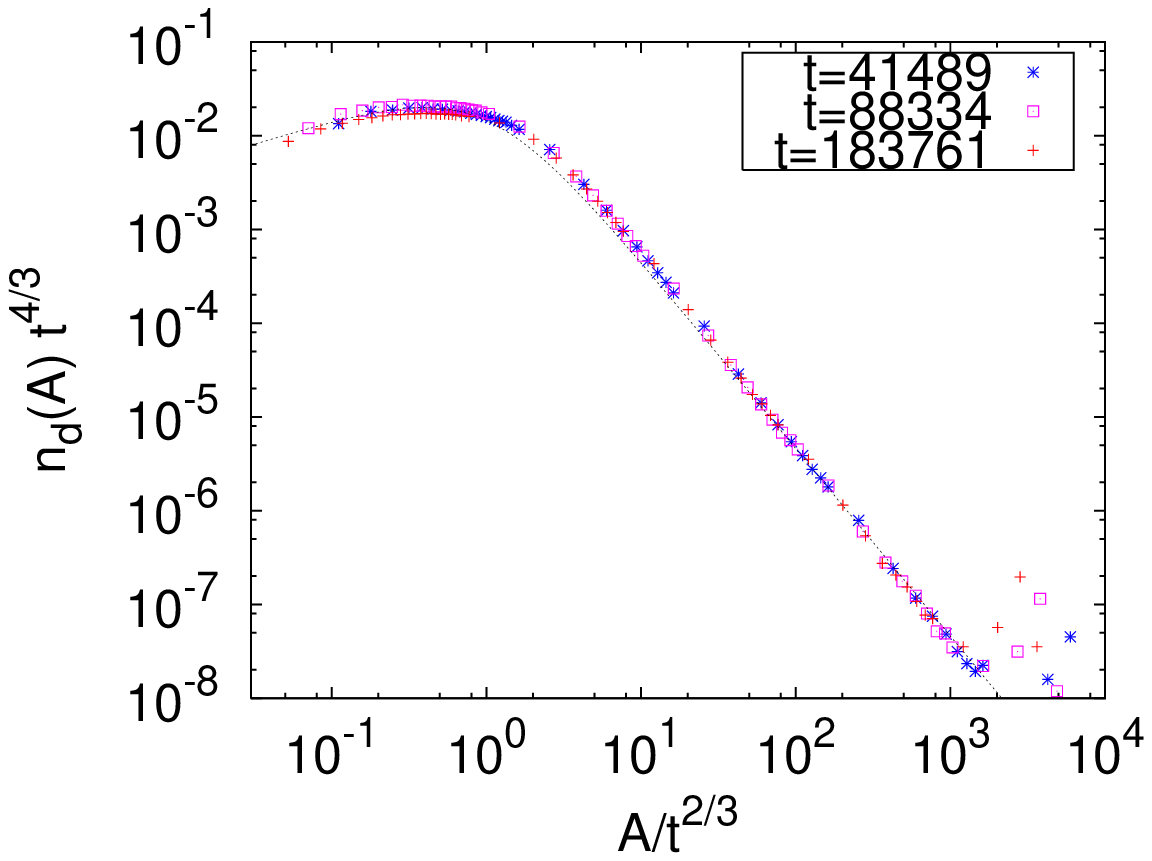}
\caption{(Colour online.)  Top panel: Number density of domains areas per unit
  area for the $2d$IM evolving at $T=1.0$ after a quench from
  $T_0\to\infty$ using the accelerated algorithm.  The dotted line
  represent the analytical prediction.  Bottom panel: rescaled data
  using the typical domain area time-dependence ${\rm A} \sim
  t^{2/3}$. The dotted line is the theoretical prediction in
  Eq.~(\ref{guess1confirmed}) with $\tau$ replaced by $\tau'$, appropriate 
  to a disordered initial state.}
\label{fig:domains}
\end{figure}

In Fig.~\ref{fig:domains} (top) we show the time-dependent domain area
distribution in double logarithmic scale, at three different times,
following a quench from $T_0\to\infty$ to $T=1.0$.  The working
temperature is very low compared to the critical value, $T_c=2.269$.

The figure shows a strong time
dependence at small areas and a very weak one in the tail, which is
clearly very close to a power law. The curves at small areas move
downwards and the breaking point from the asymptotic power law decay
moves towards larger values of $A$ for increasing $t$.  We include the
spanning domain in the statistics: the bumps on the tail of the
distribution is a finite size effect visible only when the number of
domain areas has already decreased by several orders of magnitude. In
the tail of the probability distribution function (pdf) the numerical
error is smaller than the size of the data points.  The discussion
of finite size effects in \cite{Alberto-PRE} also applies to this case.

\begin{figure}[h]
\includegraphics[width=225pt]{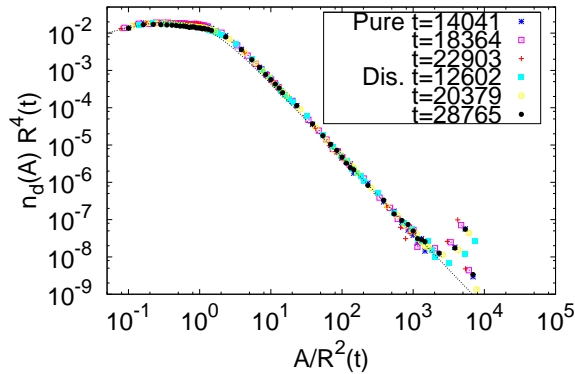}
\caption{(Colour online.) 
Simulations with the Kawasaki algorithm, without and
  with disorder. The solid lines represent the analytic
  prediction with $c_d=0.025$, $\tau'=2.055$ and $\lambda_d= 0.3$.
  For the disordered case, the growth law ${\rm R}(t)$ is extracted
  from the spatial correlation function. Even if there is deviations 
  from the theoretical curve, the data seems to show super-universality.}
\label{fig:kawa}
\end{figure}

We test the analytic prediction: the very good agreement between the
analytical theory and the data is quite impressive.
In Fig.~\ref{fig:domains} (bottom) we scale the data by plotting
$(\lambda_d t)^{4/3} n_d(A,t)$ against $A/(\lambda_d t)^{2/3}$ with
$\lambda_d=0.0083$.  For $A$ larger than the `typical' value $(\lambda_d
t)^{2/3}$ the time and $\lambda_d$ dependence become less and less
important.  We fit the parameter $\lambda_d(T)$ by analysing the
behaviour at small areas, $A^{3/2}<\lambda_d(T) t$, and we find, that
$\lambda_d(T=1.0)=0.0083$ yields the best collapse of data.  We use the
value $c_d=0.025$~\cite{Alberto-PRE}.  The full line is our prediction
Eq.~(\ref{guess1confirmed}).

In Fig.~\ref{fig:kawa} we present the domain area distribution
for the evolution using the Kawasaki algorithm both for the clean and
the disordered system. The agreement with the
analytical prediction is as good as with the bulk-diffusion algorithm,
suggesting the validity of super universality between both dynamics.

We extract the
growing length ${\rm R}(t)$ from the analysis of the spatial
correlation function~\cite{Alberto-PRE} and we find very good agreement between the
numerical data and the scaling function suggesting that
super-universality with respect to the inclusion of disorder in the
interactions also holds.

\subsubsection{Critical temperature initial condition, low $T$ evolution}
\label{subsubsec:Tc}

\begin{figure}[h]
\includegraphics[width=225pt]{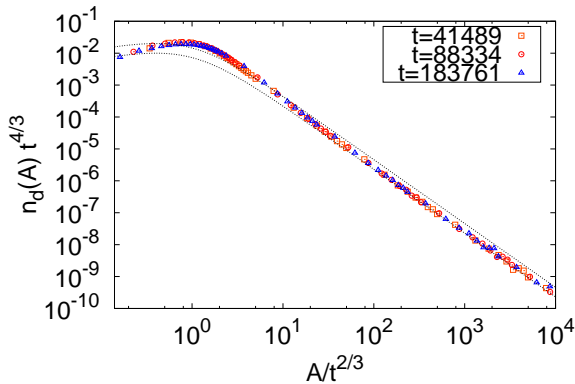}
\caption{(Colour online.)  Number density of domain areas per unit
  area for the $2d$IM evolving at $T=1.0$ after a quench from
  $T_0=T_c$ using the accelerated bulk algorithm. The lines represent 
  Eq.~(\ref{guess1confirmed}) with constant $c_d$ and $2c_d$.}
\label{fig:domains-Tc}
\end{figure}

We now use a critical Ising initial condition, $T_0=T_c$, and show the
results in Fig.~\ref{fig:domains-Tc}.  In this case we have to further
distinguish large and small domain sizes. We find that the
distribution at large areas, $A\gg (\lambda_d t)^{2/3}$ is well
described by the initial condition form.  At
the crossover $A\sim (\lambda_d t)^{2/3}$ the data points leave the
asymptotic power-law with parameter $c_d$ to approach the one with
$2c_d$. For small areas, $A <(\lambda_d t)^{2/3}$ we find that
Eq.~(\ref{guess1confirmed}) is not satisfied, see
Fig.~\ref{fig:domains-Tc}. Although few data points fall in this
regime the departure from $\sqrt{x}$ is clear.

We believe that the reason why the analytic prediction for 
small areas fails in this case is that the assumption of 
independence of small domains is not justified 
for a critical Ising initial condition.

\subsubsection{Dependence on the working temperature}
\label{subsubsec:T}

\begin{figure}[h]
\includegraphics[width=225pt]{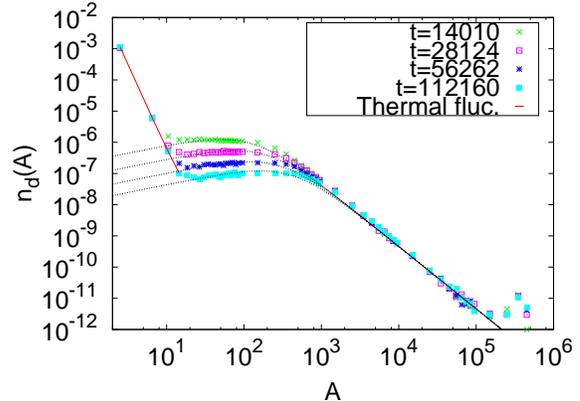}
\caption{(Colour online.)  Number density of domains areas per unit
  area for the $2d$IM evolving at $T=1.5$ after a quench from
  $T_0\to\infty$ using the accelerated bulk algorithm. The prefactor
  $\lambda_d(T)$ in the growing-length is chosen to be
  $\lambda_d(T)=0.0060$.  Compare to the results shown in
  Fig.~\ref{fig:domains} obtained for a lower
  working temperature. }
\label{fig:domains-T}
\end{figure}

Up  to now  we  showed results  obtained  using a  rather low  working
temperature. We  now study  whether and how  our results  are modified
when using higher values  of $T$. Figure~\ref{fig:domains-T} shows the
numerical data  and proves that  the scaling is well-satisfied  at all
times  and  for   all  areas.   The  large  scale   behaviour  of  the
distribution is not modified by $T$ and all data are well-described by
the  initial  condition  form.  Instead,  the
small    scale    behaviour    depends   strongly    on    temperature
fluctuations.  The anomalous  up-rising  part of  the distribution  at
small areas
is  time-independent,  suggesting  that  it  can  be  associated  with
equilibrium   fluctuations   of   the   domain-walls.  As   shown   in
Fig.~\ref{fig:domains-T},  it  is possible  to  extract the  interface
thermal fluctuations as was done in ref.~\cite{Alberto-PRE} with the
bulk equilibrium droplets (see Figs. 20 and 21 of that paper).

\section{Statistics of perimeters and fractal properties} 
\label{sec:domain-walls}

The analytic argument described in Sect.~\ref{sec:analytic} can be
extended to study the distribution of domain wall lengths or
perimeters. The domain perimeter is the total length of the interface
between the chosen domain and the neighbouring ones -- including the
hull and internal borders.  In this Section we present the analytic
prediction for this function together with numerical results that
confirm it.  We concentrate on $T_0\to\infty$ and low-working
temperature.  In the simulations we define the length of the boundary
as the number of broken bonds.

A detailed description of the relation between the domain areas and
their boundaries in the critical Ising and critical percolation
conditions as well as the number densities of domain perimeters was
given in \cite{Alberto-PRE}. Here we focus on their evolution.

\begin{figure}[h]
\begin{center}
{
\includegraphics[width=8cm]{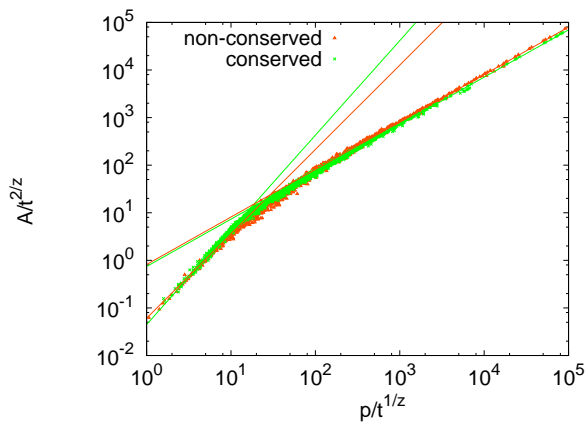}
}
\end{center}
\caption{(Colour online.)  The time-dependent relation between the
  area and the domain boundary evolving at $T=1.0$ after a quench from
  $T_0\to\infty$ using conserved and non-conserved order parameter
  dynamics. The lines are fits to the data points.  For the large
  structures the fit yields ${\alpha'}^>\simeq 1$ in both cases while
  for the small structures it yields ${\alpha'}^>\simeq 2$ for
  conserved order parameter dynamics (circular domains) and
  ${\alpha'}^>\simeq 1.8$ for non-conserved order parameter dynamics.
Here $z =2$ and 3 for non-conserved and conserved dynamics respectively.}
\label{fig:t-dep-Ap}
\end{figure}

After a quench from $T_0\to \infty$, the domain areas, $A$, and their
corresponding perimeters, $p$, obey the scaling relations (see
Fig.~\ref{fig:t-dep-Ap})
\begin{equation} 
\frac{A}{(\lambda_d t)^{2/3}} 
\sim \eta'_d 
\left(\frac{p}{(\lambda_d t)^{1/3}}\right)^{\alpha'_d},
\label{eq:t-dep-ApTc}
\end{equation}
with
\begin{equation}
\left.
\begin{array}{l}
{\alpha'}^>_d \sim 1.00 \pm 0.1
\\
{\eta'}_d^> \sim 0.75
\end{array}
\right\}
\;\; \mbox{for} \;\; 
 \frac{A}{(\lambda_d t)^{2/3}} \; \stackrel{>}{\sim} 10 \; , 
\end{equation}
and
\begin{equation}
\left.
\begin{array}{l}
{\alpha'}^<_d \sim 2.00 \pm 0.1
\\
{\eta'_d}^< \sim 0.045
\end{array}
\right\}
\;\; \mbox{for} \;\; 
 \frac{A}{(\lambda_d t)^{2/3}} \; \stackrel{<}{\sim} 10 \; . 
\end{equation}
The relation between areas and perimeters exhibits two distinct regimes
with a quite sharp crossover between them.  During the coarsening
process a characteristic scale ${\rm R}(t) \sim (\lambda_d t)^{1/3}$
develops such that domains with area $A>{\rm R}^2(t)$ have the same
exponent as in the initial condition (structures that are highly
ramified with ${\alpha'}^<\simeq 1$) and domains with $A<{\rm R}^2(t)$
become regular (${\alpha'}^< \simeq 2$). Interestingly, the small 
structures in the non-conserved order parameter dynamics 
are not completely circular, as demonstrated by the fact that 
their ${\alpha'}^<\simeq 1.8$, see Fig.~\ref{fig:t-dep-Ap}.

In analogy with the derivation in Sect.~\ref{sec:analytic} for the
time-dependent number density of domain areas, the time-dependent
number density of domain-wall lengths, $n_d(p,t)$, is given by
\begin{equation}
(\lambda_d t)
\; 
n_d(p,t) \sim 
\frac{
{\alpha'}^<_d ({\eta'}^<_d)^{3/2} \ 2c_d 
\; 
\left(\displaystyle
\frac{p}{(\lambda_d t)^{\frac{1}{3}}}\right)^{\frac{3\alpha^<_d-2}{2}}} 
{\left[ 
1+({\eta'}^<_d)^{3/2} \ \left(\displaystyle\frac{p}{(\lambda_d t)^{1/3}}
\right)^{\frac{3{\alpha'}^<_d}{2}} 
\right]^{\frac{2\tau'+1}{3}}
}
\label{eq:analytic-np-small}
\end{equation}
for small areas, $A/(\lambda_d t)^{2/3}<10$, and 
the same expression with 
${\eta'}_d^<$ and ${\alpha'}_d^<$ 
replaced by  
${\eta'}_d^>$ and ${\alpha'}_d^>$
for large areas $A/(\lambda_d t)^{2/3}>10$.  
Note that these expressions satisfy scaling.
The scaling function, $f_<(x)$, with
$x=p/(\lambda_d t)^{1/3}$, reaches a maximum at 
\begin{equation}
x_{max} = \left(\frac{3{\alpha'}_d^<-2}{2({\eta'}^<_d)^{\frac{3}{2}} [{\alpha'}_d^<(\tau'-1)+1]}\right)
^{2/(3{\alpha'}_d^<)}
\label{eq:maximum}
\end{equation} 
and then falls-off to zero as another power-law.  There is then a
maximum at a finite and positive value of $p$ as long as
${\alpha'_d}^<>1$, that is to say, in the regime of not too large
areas. The numeric evaluation of the right-hand-side yields $x_{max} =
p_{max}/(\lambda_d t)^{1/3} \sim 3$ which is in the range of validity
of the scaling function $f_<$. The time-dependent perimeter number
density for long perimeters falls-off as a power law $f_>(x) \sim
x^{{\alpha'}_d^> (1-\tau')-1}$. Although the function $f_>$ also has a
maximum, this one falls out of its range of validity.  The power law
describing the tail of the number density of long perimeters is the
same as the one characterising the initial distribution.

\begin{figure}[h]
\begin{center}
\includegraphics[width=8cm]{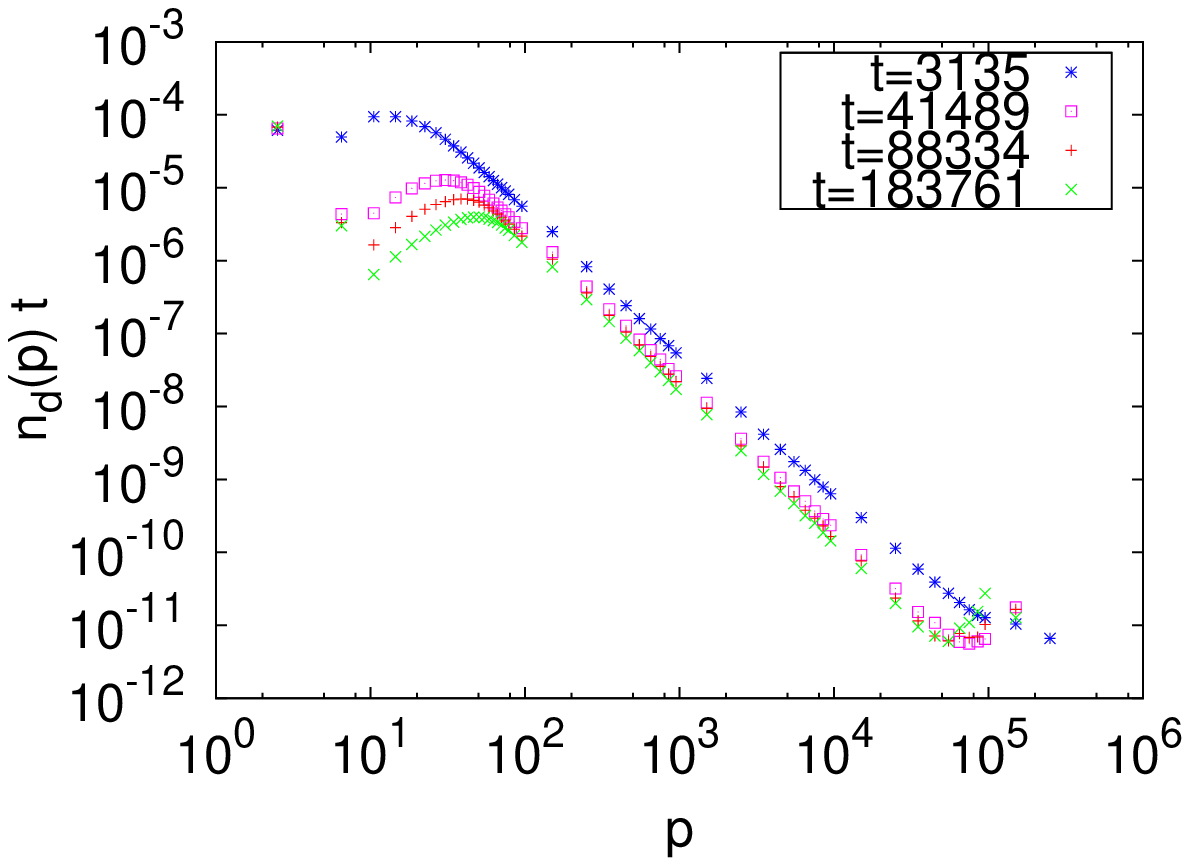}
\includegraphics[width=8cm]{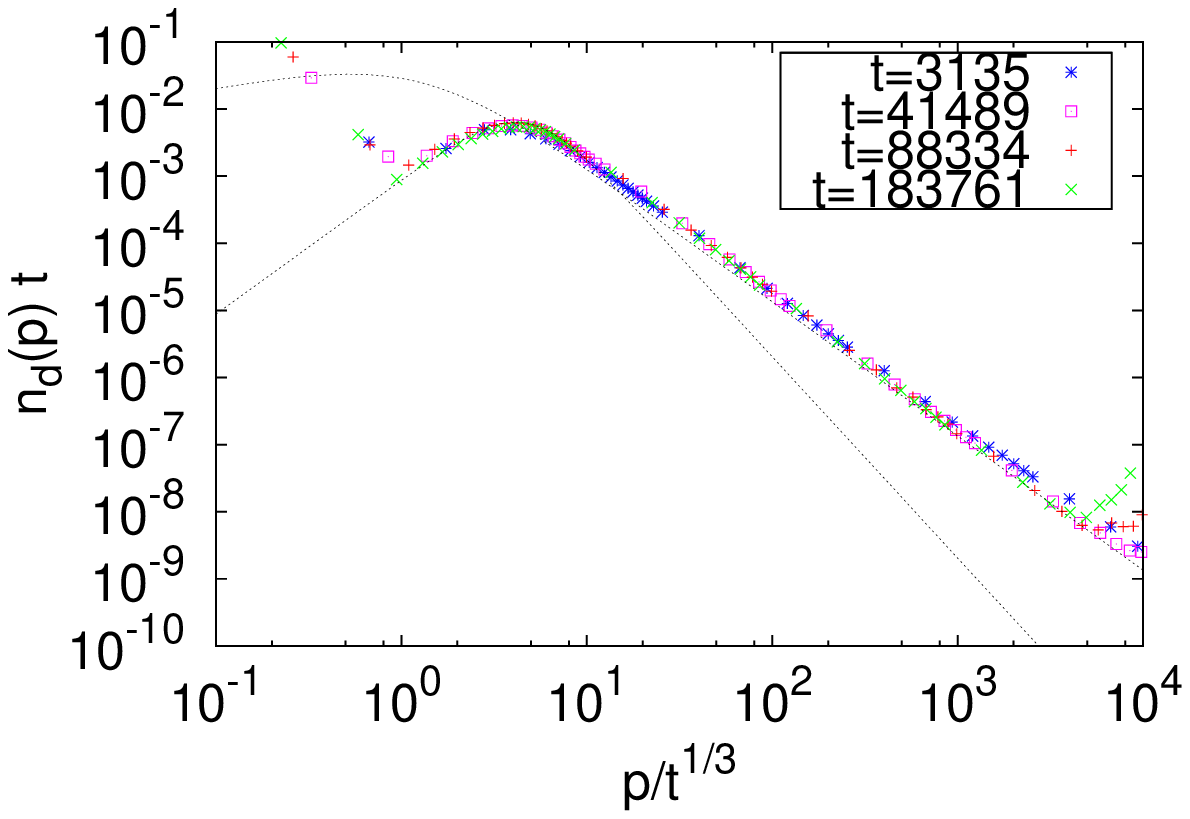}
\end{center}
\caption{(Colour online.) The time-dependent number density of
perimeters evolving at $T=1.0$ from an initial condition at 
$T_0\to\infty$. Top panel: raw data; note that the time-dependence is
visible in the whole range of values of $p$ (while in the area number 
densities the large area tails were very weakly dependent on time).
Bottom panel: scaled data and analytic predictions for the small and 
large area regimes.}
\label{fig:perimeters-hulls}
\end{figure}

In Fig.~\ref{fig:perimeters-hulls}, top and bottom, we display the
time-dependent perimeter number densities for a system evolving at
$T=1.0$ after a quench from $T_0\to\infty$.  Notice that the perimeter
length definition that we use on the lattice can only take even values
and thus when constructing the histogram we have to take into account
the extra factor of $2$ in the binning.  The data are in remarkably
good agreement with the analytic prediction; the lines represent the
theoretical functional forms for long and short lengths, and describe
very well the two limiting wings of the number density. The maximum is
located at a value that is in agreement with Eq.~(\ref{eq:maximum}).

\section{Conclusions}

In this paper we studied the statistics and geometry of hull-enclosed
and domain areas and interfaces during 
spinodal decomposition in two dimensions. 

The analytical part of our work is an extension of what we presented
in \cite{Alberto-PRE} for the non-conserved order parameter case.  The
numerical part of it deals with Monte Carlo simulations of the $2d$IM
with locally conserved magnetization. Our main results are: 
\newline
(i) We derived the scaling functions of the number density of domain
areas and perimeters with an approximate analytic argument. The
expression that we obtained has two distinct limiting regimes.  For
areas that are much smaller than the characteristic area, ${\rm R}^2(t)$, 
the Lifshitz-Slyozov-Wagner behaviour is recovered after a quench from
$T_0\to\infty$ and evolving at sufficiently low $T$. These structures
are compact with smooth boundaries, close to circular, since the
area-perimeter relation is $A\sim p^2$.  
\newline 
(ii) At higher $T$
the small area behaviour departs from the Lifshitz-Slyozov-Wagner
prediction. As for non-conserved order parameter dynamics, once we
subtracted the contribution from thermal domains within the growing
structures, the universal prediction is recovered.
\newline 
(iii) For critical Ising initial conditions large structures 
keep the geometry they had initially, but small structures fail 
to follow our prediction. We conjecture that the reason is that 
our starting assumption, independence of domain wall motion 
for small domains, is not valid due to strong correlations in this 
case.
\newline
(iv) The geometrical properties and distribution of the time-dependent 
areas that are larger than ${\rm R}^2(t)$ are the ones of
critical continuous percolation (for all initial conditions
equilibrated at $T_0>T_c$) and critical Ising (for $T_0=T_c$).  The
long interfaces retain the fractal geometry imposed by the equilibrium
initial condition.

\begin{figure}[h]
\begin{center}
\includegraphics[width=8cm]{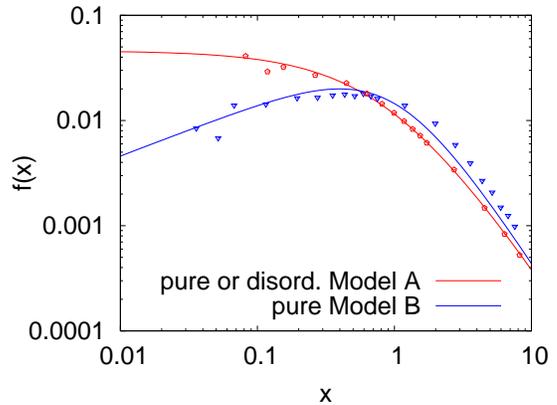}
\end{center}
\caption{(Colour online.)  Comparison between the number density of
  domain areas in the clean and disordered $2d$ Ising model with
  non-conserved dynamics and conserved dynamics in a zoom over the
  small area regime. The data points are the result of the numerical
  simulations while the straight lines are the analytic predictions. }
\label{fig:comparison}
\end{figure}

These results complement the ones that we presented in
\cite{us,Alberto-PRE,us-EPL}.  It would be interesting
to experimentally check these results, as was done for
non conserved dynamics in  \cite{us-Xtal}.
It is important to summarize
the picture that emerges from all these studies.

Large structures, with area larger than the typical one $A>{\rm R^2}$, 
keep the geometry they had initially in all respects. The scaling 
functions are thus independent of the dynamics -- conserved or 
non-conserved -- and the presence or not of weak disorder.

Small structures do not keep the initial geometry but become more
circular in all cases. For a clean system with conserved order parameter, 
domains are indeed circular while for non-conserved order parameter they 
become close to circular. The scaling function of their distribution 
depends on the type of dynamics considered but it does not depend on the 
presence of weak disorder once scaling by the pertinent growth law has been 
taken into account.  In Fig.~\ref{fig:comparison} we illustrate this
statement by comparing the number densities of domain areas, on the
one hand, in the random-bond Ising model (RBIM) with non-conserved order 
parameter (red datapoints) and the analytic result for non-conserved order 
parameter dynamics and, on the other hand, in the clean and disordered 
conserved order parameter dynamics.

Once the super-scaling of the number density with respect to weak quenched
randomness is established numerically we can infer what it implies for
the time-evolution of the individual small structures under the
assumption of independence discussed in the text. 
Indeed, starting from 
Eq.~(\ref{eq:integration}) one can derive  
Eq.~(\ref{eq:nh-allA-super}) if one assumes that each  
time-dependent area is linked to the initial value by
\begin{equation}
A^{z/2}(t) \approx {A_i}^{z/2} - {\rm R}(t)
\; ,
\label{eq:super-universality} 
\end{equation}
where $z = 2$ for systems in the universality class of model A (e.g. the  
non-conserved order parameter dynamics of the 
RBIM), and $z = 3$ for systems in the universality class of model B
(e.g. the Kawasaki dynamics). Note that inside each universality 
class, the growth laws ${\rm R}(t)$ can be very different.


\vspace{1cm}
\noindent\underline{Acknowledgements} LFC is a member of Institut
Universitaire de France. AS and LFC acknowledge financial
support from Secyt-ECOS P. A01E01 and PICS 3172, AS, JJA and LFC
acknowledge financial support from CAPES-Cofecub research grant
448/04. JJA is also partially supported by the Brazilian agencies CNPq
and FAPERGS.

\end{document}